\documentclass[traditabstract]{aa}
\usepackage{graphicx}
\usepackage{txfonts}
\usepackage{longtable}
\usepackage[authoryear]{natbib}
\bibpunct{(}{)}{;}{a}{}{,}

\def\ltsima{$\; \buildrel < \over \approx \;$}
\def\approxlt{\lower.5ex\hbox{\ltsima}}
\def\gtsima{$\; \buildrel > \over \approx \;$}
\def\approxgt{\lower.5ex\hbox{\gtsima}}

\def\mir{$\frac{B}{B+R}$}

\def\aj{\rm AJ}
\def\apj{\rm ApJ}
\def\aap{\rm A\&A}
\def\mnras{\rm MNRAS}
\def\araa{\rm ARA\&A}
\def\aar{\rm A\&AR}

\def\apjs{\rm ApJS}

\begin{document}

   \title{The horizontal branch morphology of M31 
   globular clusters. \thanks{Based on observations made with the NASA/ESA {\em Hubble 
   Space Telescope}, obtained from the Hubble Legacy Archive, which is a collaboration 
   between the Space Telescope Science Institute (STScI/NASA), the Space Telescope 
   European Coordinating Facility (ST-ECF/ESA) and the Canadian Astronomy Data 
   Centre (CADC/NRC/CSA). STScI is operated by the Association of Universities for 
   Research in Astronomy, Inc., under NASA contract NAS 5-26555.}}
   \subtitle{Extreme second parameter effect in outer halo clusters}

   \author{S. Perina$^{1,2}$, M. Bellazzini$^1$, A. Buzzoni$^1$,
   C. Cacciari$^1$, L. Federici$^1$, F. Fusi Pecci$^1$, S. Galleti$^1$}
         
      \offprints{S. Perina}

   \institute{INAF - Osservatorio Astronomico di Bologna,
              Via Ranzani 1, 40127 Bologna, Italy\\
            \email{sibilla.perina2@unibo.it} 
            \and
            Departamento de Astronom\'ia y Astrof\'isica, Pontificia Universidad Cat\'olica de Chile, 7820436 Macul, Santiago, Chile
            }

     \authorrunning{S. Perina et al.}
   \titlerunning{The horizontal branch morphology of M31 globular clusters.}

   \date{Accepted for publication on A\&A }

\abstract{We use deep, high quality colour magnitude diagrams obtained with the Hubble Space 
Telescope to compute a simplified version of the Mironov index (SMI; \mir) to parametrize the horizontal 
branch (HB) morphology for 23 globular clusters in the M31 galaxy (Sample~A), all located in the outer halo 
at projected distances between 10~kpc and 100~kpc. 
This allows us to compare them with their Galactic counterparts, for which we estimated the SMI 
exactly in the same way, in the SMI vs. [Fe/H]
 plane. We find that the majority of the considered 
M31 clusters lie in a significantly different locus, in this plane, with respect to Galactic 
clusters lying at any distance from the center of the Milky Way. In particular they have redder 
HB morphologies at a given metallicity, or, in other words, clusters with the same SMI value 
are $\approx 0.4$~dex more metal rich in the Milky Way than in M31. 
We discuss the possible origin of this difference and we conclude that the most likely  explanation is that many globular clusters in the outer halo of M31 formed 
$\approx 1-2$~Gyr later than their counterparts in the outer halo of the Milky Way, while differences in the cluster-to-cluster distribution of He abundance of individual stars may also play a role. The analysis of another sample of 25 bright M31 clusters (eighteen of them with $M_V\le -9.0$, Sample~B), whose SMI estimates are much more uncertain as they are computed on shallow colour magnitude diagrams, suggests that extended blue HB tails can be relatively frequent among the most massive M31 globular clusters, possibly hinting at the presence of multiple populations.}


\maketitle
%

\section{Introduction}
\label{intro}

Globular clusters (GCs) are well studied and widely used tracers of the earliest epoch of galaxy
formation. They are found in any kind of galaxy, they are bright and (typically) several Gyr old, at
least in the local Universe. Probably the most useful property of GCs is that valuable estimates of
their fundamental physical characteristics, notably {\em metallicity} and {\em age}, can be obtained
with various techniques \citep[see][for a recent review and references]{bs06}. Integrated colours and
spectra can be used to study unresolved clusters in distant galaxies. In the range of distances in
which GCs can be resolved into individual stars, colour magnitude diagrams (CMD) can be derived and the
most reliable and precise estimates of the cluster parameters can be obtained from observables defined
on the CMD. 

This is especially true for cluster age, as the luminosity of the main sequence turn off (MSTO) point
is unrivaled as an age indicator \citep{rfp,gza}. The advent of large telescopes and modern CCD cameras
brought the MSTO of all Galactic GCs within reach in the last two decades, thus allowing the
establishment of a robust and accurate age scale \citep[see][for a state-of-the-art analysis and
references]{dot10}. Before  this epoch, the key age indicator which was adopted to obtain global
constraints on the Galactic halo formation from GCs was the horizontal branch \citep[HB,][]{sz78,z80}.
Low-mass stars ($m\la 1~M_{\sun}$) populate the HB in the evolutionary phase of core He burning. 
The actual temperature and luminosity of a given HB star at the beginning of this phase
(Zero Age horizontal branch, ZAHB) is determined by the complex interaction of several different
factors \citep{bob}. The overall metallicity is generally recognized as the most important parameter, but age, 
He abundance, stellar rotation, and any parameter affecting the mass loss along the red giant branch 
(RGB), 
may have a substantial role \citep{fp93,marcio09}. The realization that metallicity alone was not
sufficient to account for the complex behaviour of HB morphologies in Galactic GCs led to the so
called second parameter problem, that dominated the scientific debate in the field of Galactic
astronomy for a couple of decades \citep[see, e.g.,][and references
therein]{sw67,z80,prest,fp93,lee94,mar94,mar95,fpb,alehb,dot10,ghb}. The most recent and thorough
analyses concluded that no less than three parameters are needed to account for the HB morphology of
Galactic GCs \citep[an idea already proposed in the past, see][]{fp93,alehb}, namely 
metallicity, age, and He abundance \citep{ghb}. 

The role of He abundance should be considered
within the framework of the co-existence of multiple populations in globulars \citep[see][for a recent review and references]{grat12}.  In the most generally accepted view of the early
evolution of GCs, subsequent generations of stars have undergone
processing of the the light elements, indicative of hot H-burning via the CNO
cycle, which includes the enrichment of He \citep[see, e.g.,][for discussion and references]{euge}.
This should lead to strong effects on the HB morphology of present-day clusters \citep[see, e.g.][]{n2808,c6441,y6441,ghb}.

\begin{figure*}
\includegraphics[width=8cm]{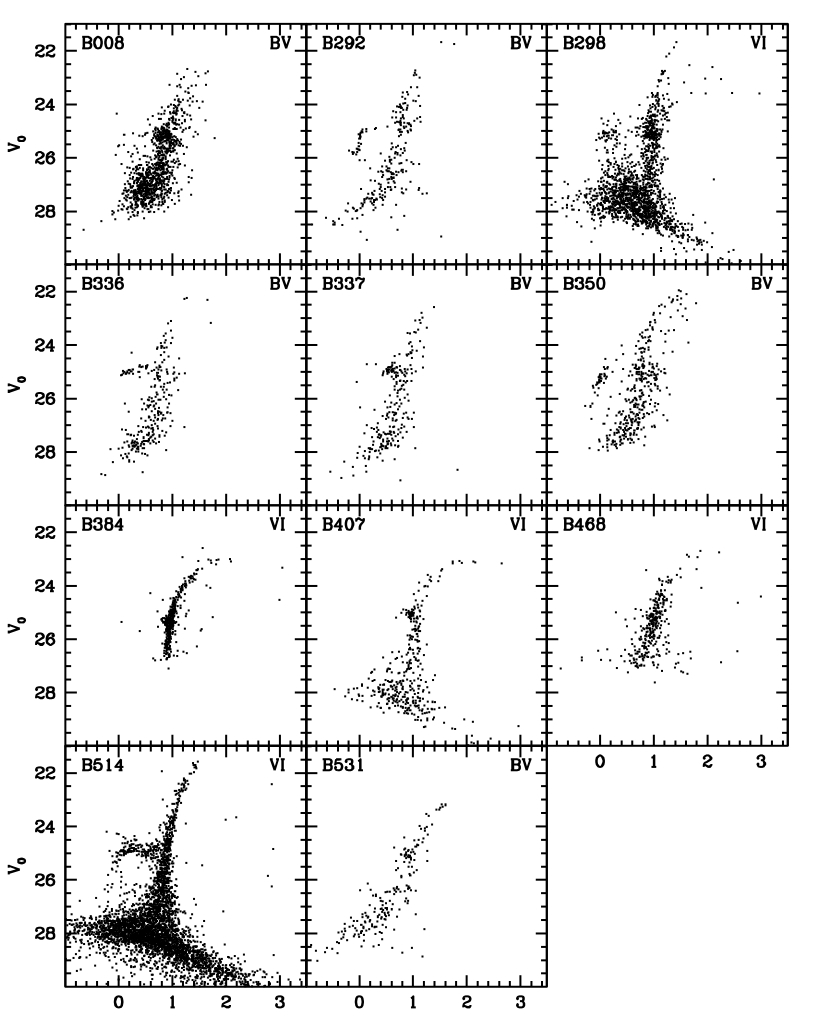}
\includegraphics[width=8cm]{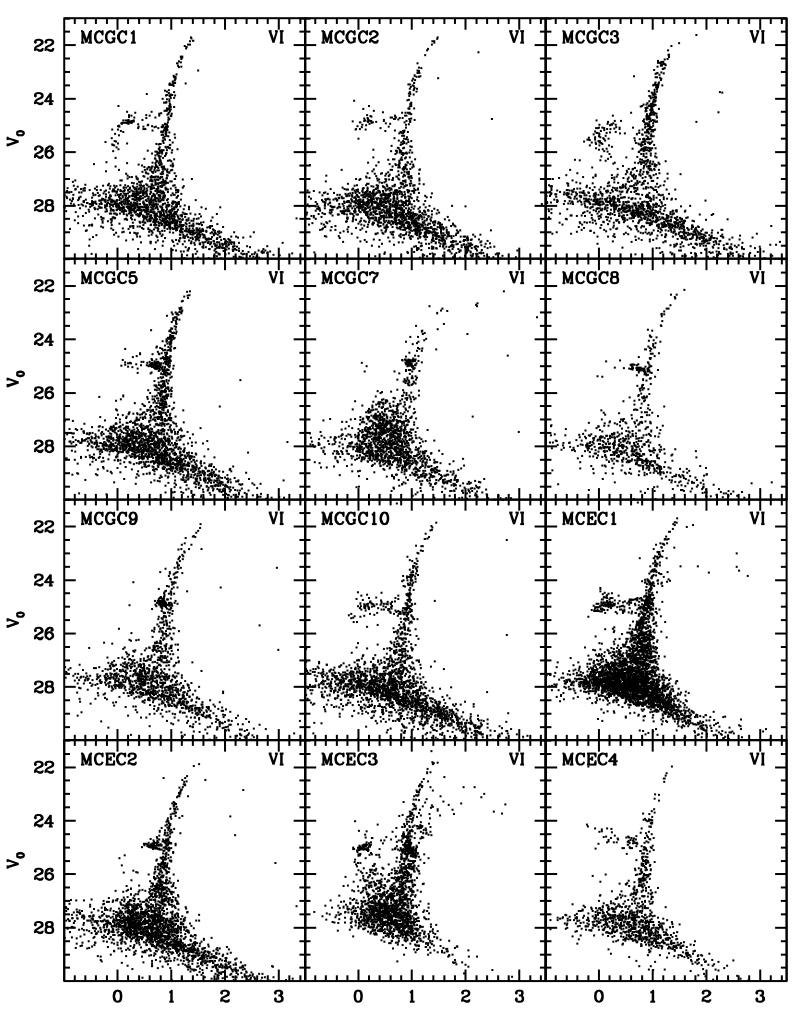}
\caption{colour-magnitude diagrams of the target clusters.
 The colour on the x axis is indicated by a label in the top-right corner of each panel : 
 BV corresponds to the dereddened colour $(B-V)_0$ and VI corresponds to 
 the dereddened colour $(V-I)_0$.}
 \label{f:cmd1}
\end{figure*}

While the use of the HB morphology as a possible \citep[albeit ambiguos,][]{fpb} age indicator  has
been superseded by much more reliable techniques for nearby clusters, it may still be valuable for more
distant systems, where the MSTO of GCs is too faint to be observed with the instrumentation currently
available. If we consider the case of our nearest neighbour giant galaxy M31, the available
observational material on GCs is quite similar to what was available for Galactic GCs in the seventies
/ early eighties, which led to the very influential scenario proposed by \citet{sz78}, i.e.:

\begin{itemize}

\item CMDs reaching down to a few magnitudes below the HB level are available for several GCs, thanks
to Hubble Space Telecope (HST) observations \citep[see, e.g.][and references
therein]{aja96,fp96,hol97,rich05,M07,per11}. It is possible to reach the MSTO luminosity of M31
clusters, but this requires such a large amount of HST time as to make a systematic study
unpractical \citep{bro04}.

\item Global metallicity estimates from integrated spectra and/or from the colour of the RGB are also
available \citep{gal09,cal11,per09}, analogously to the MW GC compilation by \citet{zw84}.

\end{itemize}

In \citet[][Paper-I hereafter]{papI} we collected a dataset of 48 M31 GCs having V,V-I or V,B-V CMD
homogeneously derived by us from images collected by different groups\footnote{The sample includes also
four objects classified as  Extended Clusters (EC), since they appear old and metal-poor as classical
globulars \citep[see][and references therein]{M06}}. In that paper, we used these CMDs to obtain an
empirical calibration of the relation between the absolute V magnitude of the HB ($M_V(HB)$) and the
metallicity ([Fe/H]), as well as a new estimate of the distance to M31. Here we use the same dataset,
and the cluster parameters homogeneously obtained in Paper-I, to get a first systematic outlook on the
behaviour of the HB morphology in the GC system of that galaxy, in particular in comparison with the
Milky Way (MW), from the distribution of individual stars in each cluster \citep[see][for an independent analysis based on integrated ultraviolet colours]{rey07,ema12}. 
We adopt a more straightforward parametrization of the HB morphology through a ``simplified'' version of the 
\citet{mir72} index MI=\mir. 
In the following we refer to this newly defined index as the Simplified Mironov Index (SMI; 
see Sect. \ref{ss:smi} for details). 

A previous use of SMI on a smaller sample of M31 GCs was made by \citet{rich05}, whereas a 
more basic approach to the HB morphology classification 
was successfully attempted by \citet{per11}, specifically aimed at verifying 
the effect of the HB morphology on the H$_{\beta}$ spectral index.  

The present analysis should be considered as a first attempt to provide a view of the M31 GC system at a comparable level of accuracy 
as we had about 30 years ago for our own Galaxy (but with different selection biases, see Sect.~\ref{bias}),
which is a remarkable step forward in the understanding of the formation and early evolution of
M31.  In Sect.~\ref{sample} we present our sample and assumptions, and we describe the procedure to
obtain the HB morphology indicator used in the analysis. In Sect.~\ref{results} we compare the
HB morphology of the considered M31 clusters with that of MW GCs, and discuss the possible origins of
the observed differences. Finally, in Sect.~\ref{summ}, we summarize and discuss our results 
also in the context of the literature.

\section{Data and data analysis}
\label{sample}

From the dataset presented in Paper-I we selected two samples of clusters for our analysis, 
depending on the quality of the available CMD:

\begin{itemize}

\item Sample~A: 23 clusters  with well defined CMD sequences and easily identifiable HB, limiting
magnitude $V_0\ga 27.0$ (i.e. reaching approximately 2 mag below the HB level, at
least), and low contamination from the M31 field. This constitutes the bulk of our analysis.
The CMD for all the clusters in Sample~A are shown in Fig.~\ref{f:cmd1}, to give the reader a
clear idea of the quality of the HB morphology classification that can be obtained from this material. It must be noted that the quality of the data for Sample~A clusters is not fully homogeneous. The CMDs of B384 and B468 have a brighter limiting magnitude than most of the other clusters of this sample. B008, B292, B298, B336, B337, B350, and  B531 suffer from a slightly higher degree 
of contamination and/or crowding, w.r.t to the other Sample~A clusters, and/or may be affected by some differential reddening (still not seriously affecting their HB morphology).

\item Sample B: 25 clusters with CMD not fulfilling the above criteria, still valuable, however, to
obtain an estimate of SMI, albeit significantly more uncertain than for Sample~A. We briefly consider Sample B here only for completeness and to get preliminary insight for future investigations.

\end{itemize}

In Paper-I all the cluster CMDs were compared with a grid of RGB and HB templates of Galactic GCs with a
fitting procedure that provides simultaneous best estimates of the HB magnitude level $V_{HB}$, the reddening
E(B-V), the metallicity [Fe/H], and the distance modulus for each cluster (see Paper-I for a detailed
description and discussion about uncertainties; we discuss the role of these uncertainties
in our analysis in Sect.~\ref{results}). In the following we adopt $V_{HB}$, E(B-V) and [Fe/H]
from Paper-I; it is clear that Sample~A clusters, having the best CMDs, have also the most reliable and
accurate estimates of these parameters.

All the results of the present study are based on the comparison with Galactic GCs taken as a reference. As
an homogeneous Galactic sample, we adopt the publicly available HST photometry of 74 Galactic GCs from
\citet{snap}\footnote{The original data \citep{snap} were obtained in the F439W and F555W bands of the Wide Field and Planetary Camera 2 (WFPC2). 
Photometry in these filters has been converted into standard B,V photometry as described in Paper-I.}. 
The values of [Fe/H], $V_{HB}$, E(B-V) and $(m-M)_0$ for these clusters have been taken 
from \citet[][2010 version]{har96}. 
The adopted extinction laws are the same as in Paper-I, $A_V =3.1E(B-V)$, $A_I =1.94E(B-V)$ and $E(V-I)= 1.375E(B-V)$. 
The list of Sample~A and Sample~B clusters and the relevant parameters for the
present analysis are provided in Table~\ref{Tab1}.

\subsection{The Simplified Mironov Index (SMI)}
\label{ss:smi}

 In the widely used original Mironov Index MI=\mir  
\citep[see e.g.~][and references therein]{sz78,lee94,mar94}, 
R and B are the numbers of HB stars respectively redder and bluer than the RR Lyrae instability 
strip edges. Here we use the same formulation, but R and B are defined as the number of 
HB stars redder and bluer than a given colour threshold, approximately located in the middle 
of the RR Lyrae instability strip. 
Given the sometimes scanty population on the HB, this appeared as the most sensible 
choice allowing us to account for all the HB stars with a single  
parameter.   

\begin{figure}
\includegraphics[width=80mm]{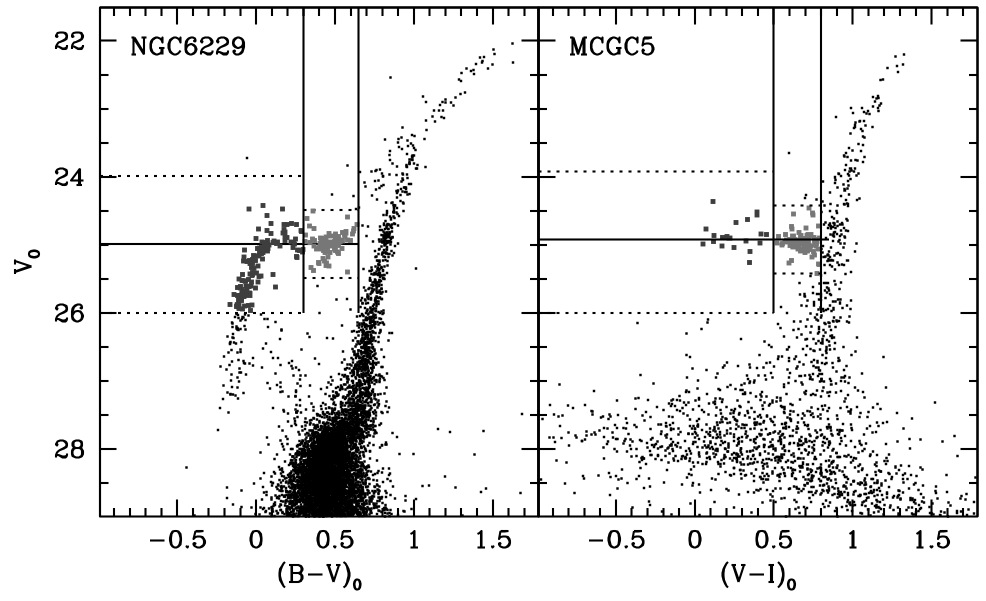}
 \caption{Illustration of the selection boxes adopted for the computation of the Simplified Mironov
Index.  Left panel: CMD of the Galactic globular cluster NGC6229, reported to the mean distance of M31. Right panel: CMD of the Sample~A globular cluster MCGC5. In each panel, the vertical segments mark, from left to right, the boundary between the blue (B)
and red (R) part of the HB and the red limit of the HB selection.
The continuous horizontal line is the mean level of the HB, while the dotted horizontal segments enclose the stars selected as bona-fide HB. In both diagrams Blue and Red HB stars selected to compute SMI are plotted in dark and light grey, respectively.}
 \label{f:method}
\end{figure}

We estimated the SMI from extinction-corrected CMDs as illustrated in Fig.~\ref{f:method} 
and described below:

\begin{enumerate}

\item For each cluster we consider a circular area around the center that is clearly dominated by
cluster members. All the star counts described below are performed in this {\em selected circle}.

\item We count as R the HB stars lying within $\pm 0.5$~mag from $V_{HB}$ and having 
$0.50<(V-I)_0\le0.80$  
or $0.30<(B-V)_0\le0.65$, 
depending on the photometric bands of the available CMD.

\item We count as B the HB  stars having $(V_{HB}-1.0)<V_0<26.0$, and $(V-I)_0\le0.50$ or 
$(B-V)_0\le0.3$, depending on the available CMDs. The $V_0<26.0$ limit has been adopted to avoid
regions of the CMD were the completeness level can be sharply falling, thus strongly biasing the star
counts. Note that {\em the same selection boxes are adopted for M31 and MW GCs}\footnote{The $V_0<26.0$
limit for the selection of B stars adopted for M31 clusters correspond to $M_V=1.58$, adopting the
average distance modulus from Paper-I, $(m-M)_0=24.42$. This has been converted into the corresponding
limit for each Galactic GC adding their true distance moduli.}. This implies that a fraction of 
genuine and clearly identifiable Blue HB (BHB) stars can be excluded in some Galactic GCs (as in the case of 
NGC6229, shown in Fig.~\ref{f:method}), to preserve the full homogeneity in the comparison with 
M31 clusters. The case of NGC6229 shows also that the same threshold helps to prevent contamination of the BHB star counts from blue stragglers. In any case, stars appearing to be associated with the Blue Straggler sequence are excluded from the computation of SMI.

\item For Sample~A clusters, the average number of field stars falling in each selection box was
estimated using an annulus ({\em field annulus}) surrounding the cluster with the same area as the 
{\em selected circle}. Field counts are subtracted from the cluster counts and SMI is computed 
propagating Poisson errors on the star counts. The effects of uncertainty in $V_{HB}$ are negligible, 
because of the generous interval in V that is considered in the selection of HB stars. 
The uncertainty in E(B-V) can have a significant impact, in principle: we consider the case in 
detail in Sect.~\ref{results}.  

\item For Sample B clusters, we directly estimate B and R from statistically decontaminated CMDs
\citep[see][and Paper-I]{per09}. This is a less robust procedure for star counts with respect to that
adopted for Sample~A, but it was the only viable one, since the identification of key CMD features 
was too uncertain without previous decontamination.

\end{enumerate}

The location of the threshold between R and B stars is arbitrary and has been set only for 
convenience. However, since the aim is the comparison between two sets of GCs where the SMI 
has been
computed in the same way, this cannot introduce any serious bias in the analysis. 
On the other hand, the consistency between the threshold in $(B-V)_0$ and $(V-I)_0$ has been 
checked for accuracy using clusters that have photometry in all the three passbands.

In a few clusters the  red HB is totally or partially superposed to the RGB. In these cases we
estimated the contribution of these stars to R by constructing the luminosity function (LF) of RGB stars in the {\em selected
circle} and in the corresponding  {\em field annulus}. Once subtracted the latter from the
former, to remove any possible contamination from field Red Clump stars, we look for residual
peaks in the decontaminated LF that  can correspond to the red HB. Great care is taken to
discriminate between the red HB and the (fairly smaller and, generally, brighter) peak produced by the RGB bump. Then the
number of stars in the red HB peak is estimated by subtracting the underlying RGB, interpolated in
the region of the red HB peak. We checked several clusters in this way for the presence of red HB
stars superposed to the RGB, and a significant signal was detected only for two clusters, MCGC8
(36 R stars) and MCEC1 (46 R stars). In some cases, like e.g. B468, although the red HB is
clearly superposed to the RGB, the above procedure is pointless as SMI vanishes all the way. For B298 
and MCEC3 the uncertainties on the number of R stars are somewhat larger because of a stronger
field contamination. For these two clusters we provide only upper limits to their SMI,
but the true value cannot be much different from the reported one.

\begin{figure}
\includegraphics[width=8cm]{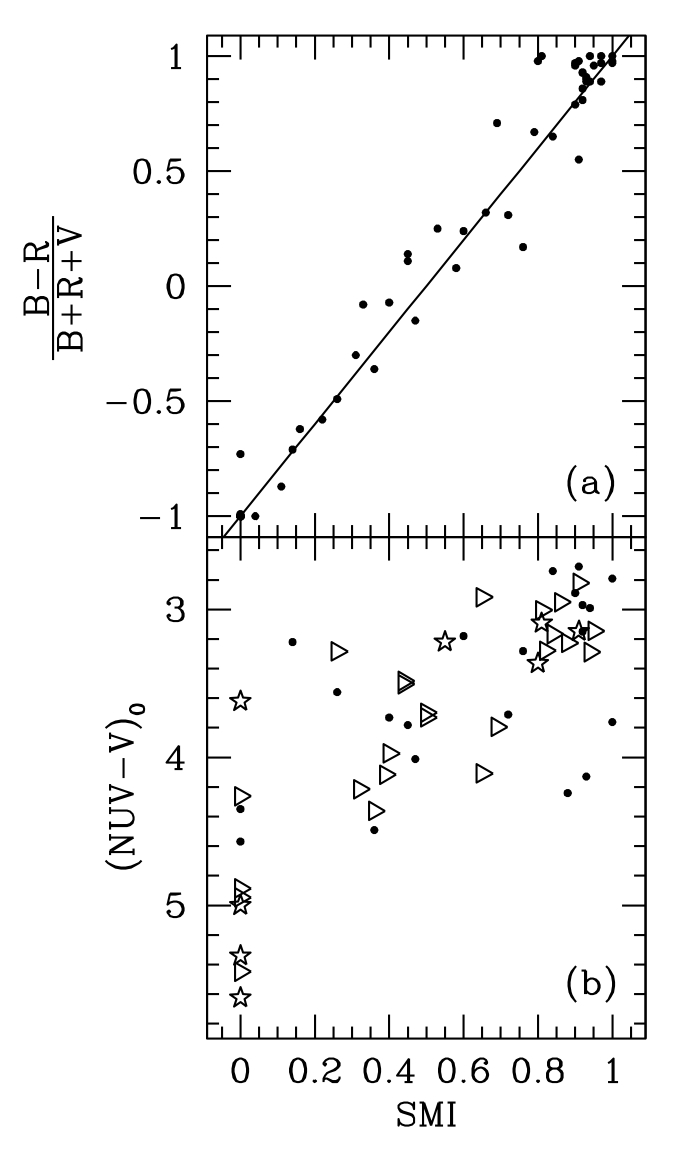}
\caption{{\em Panel (a):} Correlation of the Simplified Mironov Index with the HBR index ($\frac{B-R}{B+R+V}$) for our sample of Galactic GCs. HBR values are taken from the latest version of the \citet{har96} catalogue.
The continuous line is the relation between \mir and $\frac{B-R}{B+R+V}$ from \citet{prest}, reported in Eq.~\ref{presto}. {\em Panel (b):} Correlation of SMI with the 
integrated ultraviolet-optical colour $(NUV-V)_0$  for Sample~A (open stars), Sample~B (open triangles, to be considered as lower limits), and Galactic GCs (small filled circles).
}
 \label{f:hbrnuv}
\end{figure}

\subsection{Relation with other HB morphology indicators}
\label{HBpar}

It is clear that the SMI should have a similar behaviour as the original MI  
which, in turn, is a simpler version of the more generally used horizontal branch Ratio 
HBR=$\frac{B-R}{B+R+V}$
\citep[where V is the number of stars in the RR Lyrae instability strip,][]{lee94}. 
The two indices are tied by the relation \citep{prest}:

\begin{equation}
\label{presto}
\frac{B}{B+R}=0.50+0.50\frac{B-R}{B+R+V} 
\end{equation}

Fig.~\ref{f:hbrnuv}a shows that the same relation provides an excellent fit also to the
trend between SMI and HBR.

These indices trace essentially the {\em peak} (or the weighted mean) 
of the HB colour distribution, this being especially true for the SMI because the 
$V_0<26.0$ limit makes it blind to stars in the extreme blue tail of the HB.
Other parameters - e.g.~ L$_t$ or T$^{max}_{eff}$(HB) - are better suited to trace 
the {\em extension} of the HB, in particular of the blue tail, when present
\citep{prest,fp93,alehb,ghb}. 

It has been shown \citep{rey07,ema12} that the integrated near/far ultraviolet
- visible colours  $(NUV-V)_0$ and $(FUV-V)_0$ are good tracers of the GC HB morphology.
In principle they should be preferred to our SMI, because by definition they cannot be
affected by the incompleteness problem that forced us to adopt the $V_0<26.0$ limit, 
and so are particularly sensitive (especially $(FUV-V)_0$) to the blue tails of the 
HB distributions \citep{ema12}.  
However, these integrated colours are much more affected by reddening and contamination 
from hot non-HB stars (e.g. from the M31 disk field population or the turn-off stars 
of the GC itself, especially $(NUV-V)_0$; see, e.g., Sect.~\ref{smifuv}). 
Therefore,  SMI should provide a complementary view of the HB 
morphology, with respect to the UV-optical integrated colours.  

In Fig.~\ref{f:hbrnuv}b we show that indeed SMI and $(NUV-V)_0$  broadly anti-correlate, and the behaviour of Galactic and M31 clusters from our samples are very similar in this plane\footnote{Integrated FUV and NUV magnitudes (from {\em Galaxy Evolution Explorer - GALEX} photometry) are taken from \citet{rey07}, for M31 clusters, and from \citet{ema12} for MW clusters. The values of $(NUV-V)_0$ and $(FUV-V)_0$ have been computed 
adopting $A_{NUV}$=8.90E(B-V) and $A_{FUV}$=8.16 \citep[from][]{rey07}, and taking V magnitudes 
from RBC~V4.0 and E(B-V) from Paper-I. The adoption of the very recent updated and extended dataset by \citet{kang} would not lead to any significant improvement, for the sample of M31 clusters considered here.}.
We note that most of the M31 clusters in our samples having available NUV and FUV magnitudes are from Sample~B. It is reassuring to note that, in spite of the large uncertainties associated to the SMI estimates for Sample~B clusters, SMI anti-correlates quite well with $(NUV-V)_0$. 
The same degree of correlation is observed between $(NUV-V)_0$ and HBR \citep[see e.g.][]{ema12}. 

Given the results illustrated in Fig.~\ref{f:hbrnuv}, we consider that our SMI is validated as a reliable HB morphology indicator tracing the peak of the HB distribution as MI and HBR. 
For most of the clusters in Sample~A which lack UV-optical colours, this is the only 
parametrization of the HB morphology currently available.


\begin{figure*}
\begin{center}
\includegraphics[width=16cm]{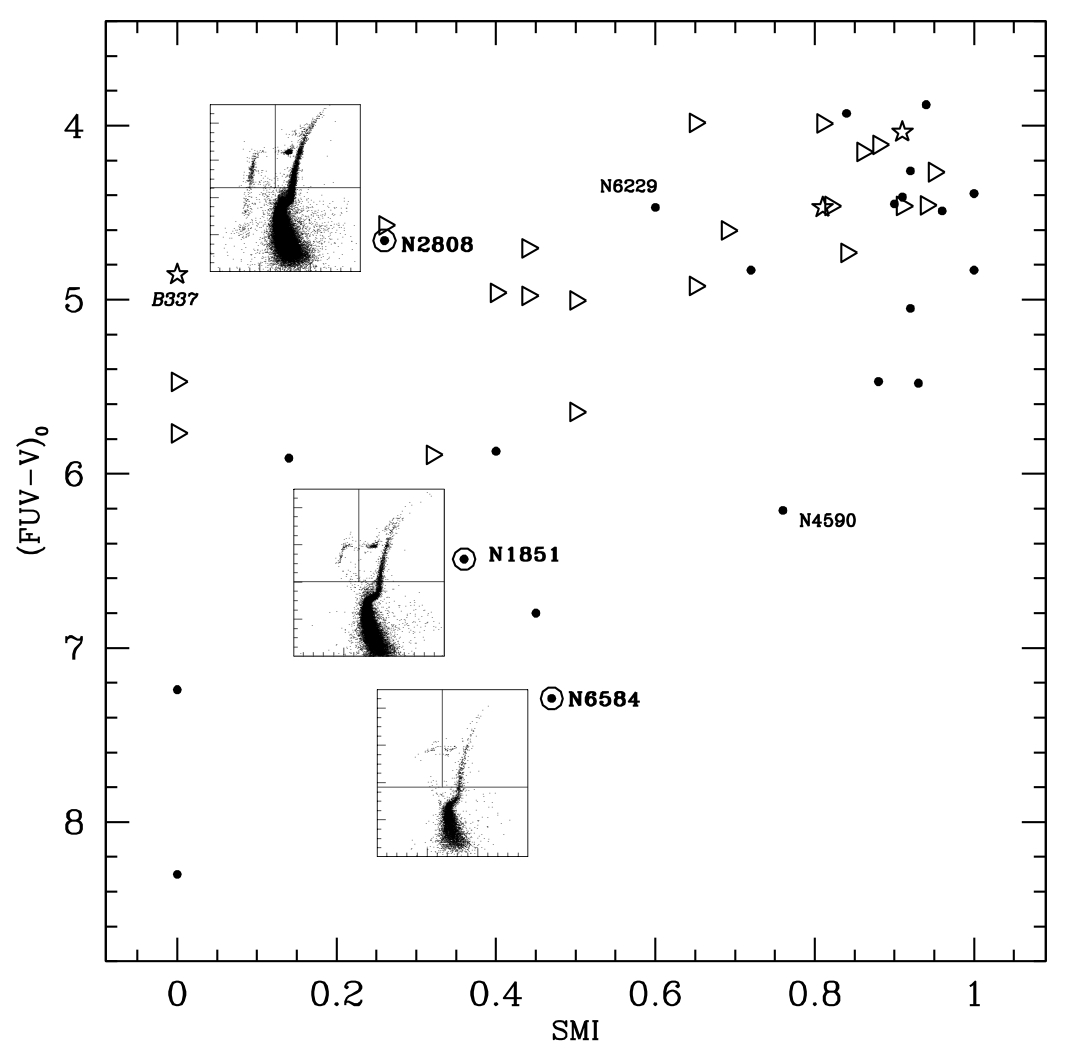}
\caption{Simplified Mironov Index versus the integrated ultraviolet-optical 
colours $(FUV-V)_0$  for Sample~A (open stars), Sample~B (open triangles, to be considered as lower limits) and Galactic GC (small filled circles). Three remarkable Galactic GCs are highlighted with a concentric open circle, are labelled in boldface and have their CMD presented in an inset box, to show the actual variation of their HB morphology with decreasing $(FUV-V)_0$ at nearly constant SMI.
Two other Galactic clusters and one Sample~A cluster, that are mentioned in the text, are also labelled.
}
\end{center}
\label{barabba}
\end{figure*}

\subsubsection{SMI vs. $(FUV-V)_0$: the effect of blue tails}
\label{smifuv}

The comparison between SMI and $(FUV-V)_0$, shown in Fig.~4, deserves a deeper discussion. The anti-correlation is significantly less tight than in the SMI vs $(NUV-V)_0$
plane and, above all, the distribution of M31 clusters is different from their MW counterparts.
In particular they are all confined in a relatively narrow strip along the blue edge of the $(FUV-V)_0$ distribution.

This is partially due to the selection in FUV flux that affects the M31 sample. At that distance clusters with FUV fluxes below a certain threshold were not detected by GALEX: in practice M31 GCs with $M_V\ga -7.0$ do not have valid measures of the integrated FUV magnitude \citep{rey07}. This translates into a threshold in $(FUV-V)_0$ colour ranging from $(FUV-V)_0\la 7.2$ for the reddest clusters (having integrated $(B-V)_0\approx 1.0$) to $(FUV-V)_0\la 5.0$ for the bluest ones (having integrated $(B-V)_0\approx 0.3$), effectively squeezing the M31 GCs in our sample into a narrower range of $(FUV-V)_0$ with respect to MW clusters.

However, there are two additional effects that can concur to produce the observed difference. First, SMI estimates for 
Sample~B clusters (the large majority of M31 clusters in Fig.~4) should be considered as lower limits, because the bright limiting magnitude of their CMD may wipe out most (or all) of their BHB stars (if present) from the computation of SMI. Larger SMI values for some of these clusters would help to distribute them over the same range covered by Galactic GCs in this plane. Furthermore, some of them may have very extended blue tails, including a population of extreme HB (EHB) and/or (possibly) {\em blue hook} stars \citep[see][for references and discussion]{marcio09}.

The inset CMD for three Galactic clusters (NGC6584, NGC1851 and NGC2808) having similar SMI and widely different $(FUV-V)_0$ illustrates this possibility and shows very clearly the complementarity of SMI and $(FUV-V)_0$ in describing the HB morphology. All the three clusters have a bimodal HB morphology, however they greatly differ in the {\em extension of their blue tail}. The hottest temperature reached by their HB corresponds to the largest integrated FUV flux and, consequently, the bluest $(FUV-V)_0$ colour \citep[see][]{ema12}. The difference in the morphology between NGC4590 and NGC6229 (also labelled in the figure and discussed in Sect.~\ref{results}, below) is similar to that between NGC6584 and NGC1851/NGC2808. 

The trend shown in the CMDs indicates that a remarkably extended BHB is required to reach the bluest limits of the $(FUV-V)_0$ range, thus suggesting that several sample~B clusters likely have a significant blue tail population that went undetected in their shallow CMDs.
{\em Hence extreme HB morphologies like that displayed by NGC2808, and a few other peculiar clusters in the Milky Way, i.e. $\omega$~Cen, M54 and NGC2419, may be relatively common among Sample~B clusters} \citep[see also][]{ema12}. Interestingly enough, such extreme $(FUV-V)_0$ colors indicate that a substantial 
fraction \citep[3-6\%,][]{buz12} of the total (bolometric) 
luminosity for these clusters is emitted shortward of 2500 \AA,
a feature that may closely deal, on a larger galactic scale, with the striking 
phenomenon of the UV upturn, as extensively observed among elliptical galaxies
\citep{yi04}.

The case of B337 is especially interesting since it has SMI=0.0, i.e. purely red HB as seen from its CMD, and extremely blue $(FUV-V)_0$ colour. Still, it is a sample~A cluster and in Fig.~\ref{f:cmd1} it can be appreciated that no significant BHB population emerges in the CMD down to $\approx 3$~mag below the red HB level, a range containing the majority of BHB stars in NGC2808. A detailed inspection of the available HST images revealed the presence of two possible contaminating sources lying within the aperture adopted by \citet{rey07} for their NUV/FUV GALEX photometry ($r_{ap}=4.5\arcsec$). A star nearly two magnitudes brighter than the brightest cluster star is located at $\approx 3.5\arcsec$ from the cluster center. Its colour ($(B-V)_0=0.45$) and magnitude are typical of a foreground Galactic dwarf. While this bright star is bluer than most of the cluster giants and clearly contaminates the integrated FUV magnitude of the cluster, it is too red to push the integrated $(FUV-V)_0$ to the extreme value observed. Another potential contaminant has been identified at $\approx 0.25\arcsec$ from the cluster center, a region where photometry packages are unable to reliably resolve and estimate fluxes of any star. A very rough estimate based on the comparison of the intensity peak with stars of known magnitudes suggest that this star is  brighter than the cluster RGB tip and may have a color as blue as $(B-V)_0\approx 0.0$. This would be consistent with a Post Asymptotic Giant Branch star \citep{jasni}, that may well dominate the UV flux of the cluster.
Hence, B337 may be the example of a cluster whose integrated UV-optical colour is strongly affected by fore/back-ground contamination and/or by a rare non-HB source. However, these cases should be quite unusual in our samples.

\subsection{An obvious selection bias}
\label{bias}

Sample~A is clearly not representative of the whole M31 GC system. The quality criteria imposed on
the CMD are very hard to fulfill for clusters that are projected on the high surface brightness bulge
or disk of M31, since there (a) the background level and the crowding are high, preventing
accurate photometry of individual stars down to the required faint magnitude limit, and (b) 
the contamination
by field M31 stars can be also very high, strongly affecting star counts on the HB. Moreover,clusters more distant from the center of their parent galaxy have intrinsically larger sizes \citep{vdb94}, hence remote M31 clusters are less severely crowded than those near the center, in average, when seen from the Earth.

\begin{figure} \includegraphics[width=80mm]{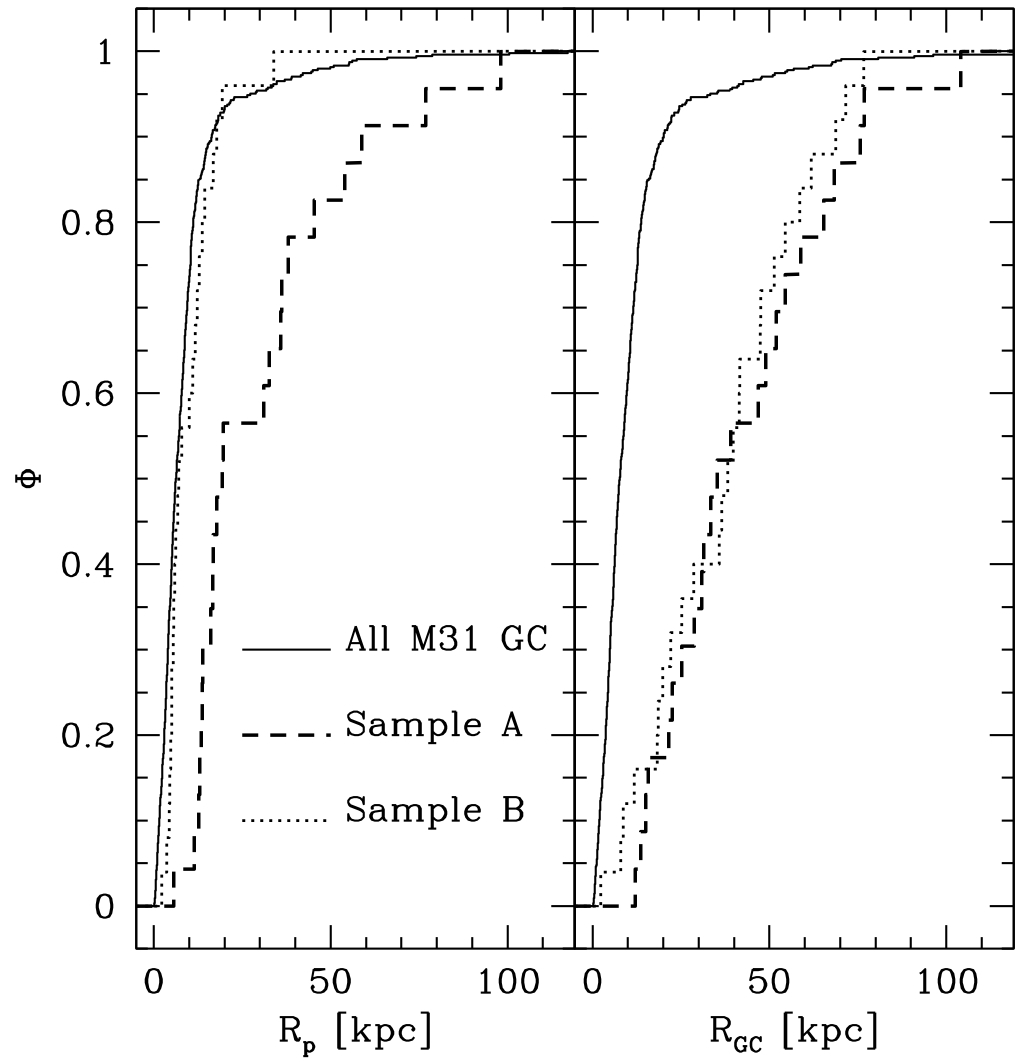} \caption{{\em Left panel:} cumulative distributions of projected distances from the center of M31 of Sample~A (long dashed line) and Sample~B (dotted line) are compared with the distribution of all the confirmed M31 clusters classified as old (GCs) (solid line). {\em Right panel:} the same comparison but for the distributions of three-dimensional distances from the center of M31 ($R_{GC}$). For the whole sample we used$\sqrt{\frac{3}{2}R_p}$ as a proxy for $R_{GC}$.}
\label{f:bias} 
\end{figure}

Indeed, the left panel of Fig.~\ref{f:bias} shows that most of Sample~A clusters lies at projected distances 
$R_p>10$~kpc from the center of M31, a radius that encloses approximately 80 per cent of the whole 
sample of M31 GCs\footnote{The data are from the Revised Bologna Catalogue of M31 globular clusters and
candidates \citet[][RBC V4.0]{gal04}; {\tt http:/www.bo.astro.it/M31/}}.  Moreover, $\approx 20$ per
cent of Sample~A clusters have $R_p>40$~kpc, a range that contains less than 5 per cent of the whole
sample. Finally, many of the outermost clusters of Sample~A have been found to be likely associated
with known existing substructures in the outer halo of M31 \citep{per09,M10}. 
{\em Therefore, Sample~A is
more likely representative of the outer halo population of the GC system of M31, and/or of the
population of GCs more recently accreted via the disruption of their parent dwarf galaxies}. 
This selection bias should be kept in mind very clearly when comparing Sample~A with
Galactic globulars. 

Sample~B shows a much more concentrated distribution of $R_p$, more similar to the overall population. However, the right panel of Fig.~\ref{f:bias} shows that the distribution of {\em spatial} (3-D) distances\footnote{Derived by combining the projected distances and the line-of-sight distances from the center of M31, the latter obtained from the individual distance moduli from Paper-I, as done also by \citet{M06,M07}.} is very similar to the one of Sample~A, and clearly also not representative of the overall M31 GC system.

It is worth recalling that \citet{M06} and \citet{M07} noted a strong second parameter effect among
the recently discovered group of remote (and/or extended) clusters of M~31 \citep{hux04,hux05,gal06}.
However, a quantitative and systematic comparison with MW clusters is attempted here for the first
time.

\section{Results}
\label{results}

\begin{figure*}
\includegraphics[width=160mm]{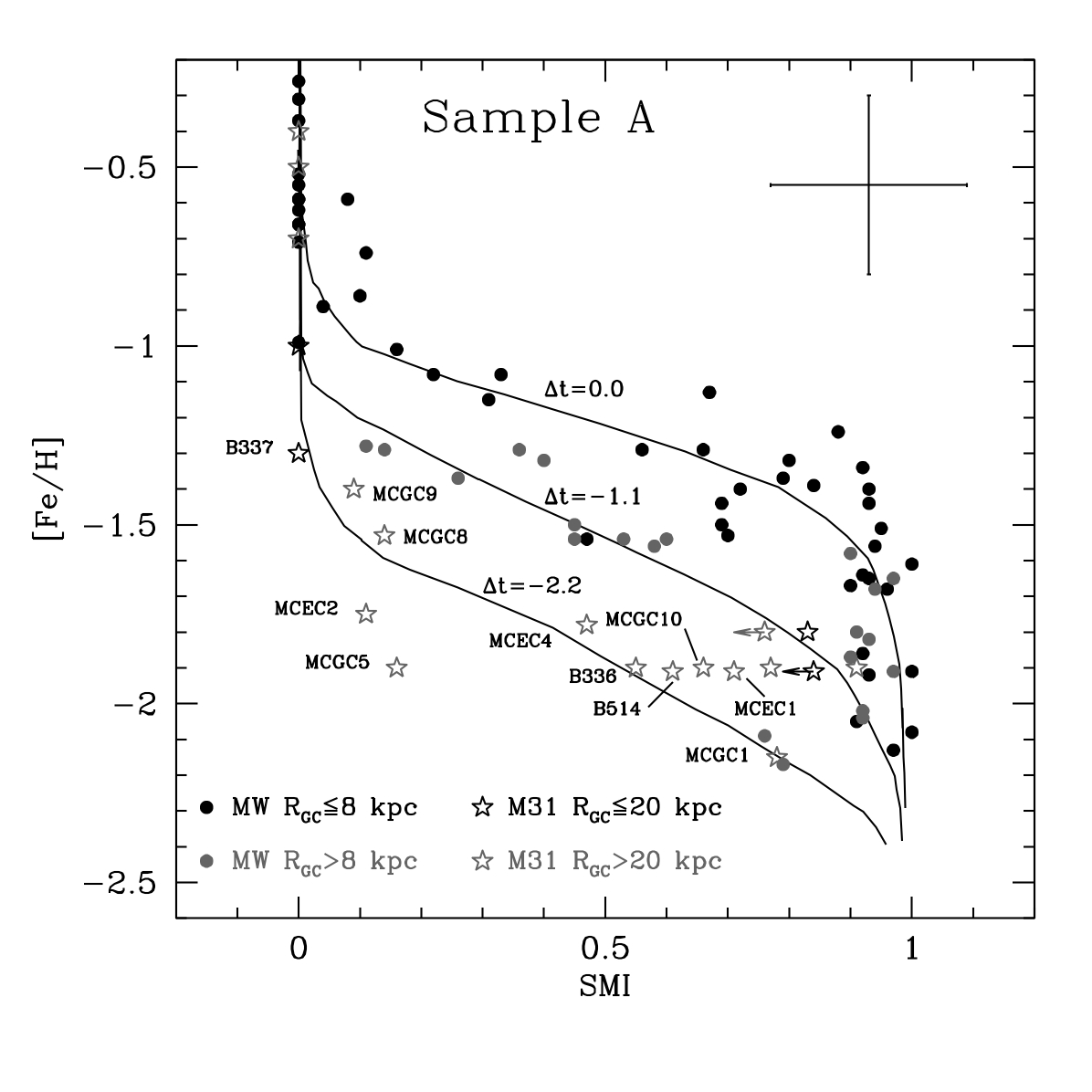} \caption{The SMI vs. metallicity diagram. The Galactic GCs are
 plotted as filled circles, the M31 Sample~A GCs as open stars.  The symbol colour code shows the  
 galacto-centric distance of the cluster, as  described in the bottom-left corner of the panel.
 The $R_{GC}=20$~kpc threshold was somehow arbitrarily adopted for Sample~A clusters to highlight those that cannot be considered as exceptionally far, since it correspond to the radius containing $\approx 90$\% of the whole population of confirmed M31 GCs (see Fig.~\ref{f:bias}).
 Arrows indicate upper limits. Solid lines are isochrones from the synthetic HB models by
 \citet{rey01}, and are labelled according to their age difference in Gyr. Typical error-bars 
 for M31 GCs are shown in the top-right corner. Sample~A clusters lying near the $\Delta t=-2.2$~Gyr 
 isochrone are labelled.}
 \label{f:mir}
\end{figure*}

The main result of the present analysis is shown in Fig.~\ref{f:mir}. Sample~A clusters are compared
with MW clusters in the classical plane opposing metallicity and HB morphology. The solid lines
show three theoretical isochrones illustrating the expected trend of the  SMI as a function of [Fe/H].
These have been obtained by converting isochrones in the $\frac{B-R}{B+R+V}$ vs. [Fe/H] plane
\citep[derived by][from synthetic HB populations models]{rey01} using  Eq.~\ref{presto}. 

It must be stressed that these isochrones are plotted only as a
general reference for the interpretation of the observed distributions, since they (a) refer to a
slightly different quantity than that plotted in Fig.~\ref{f:mir} (i.e. the original MI 
instead of SMI, but this does not seem a reason of serious concern, see Fig.~\ref{f:hbrnuv}), and (b) the details of their shape depend on the mass loss recipe adopted in
modeling the synthetic HB \citep{rey01,lee94,mar94}. The three isochrones correspond to 
different ages in steps of 1.1 Gyr, setting a rough relative age scale, from top to bottom 
$\Delta$t=0, -1.1 and -2.2 Gyr \citep[see Fig.~9 in][]{rey01}. Alternatively, they can be read 
as a relative scale of
He abundance Y, approximately corresponding to $\Delta$Y=+0.019, 0.0 and -0.019 
(M. Catelan, private
communication)\footnote{The adopted standard Y abundance is Y=0.23+2Z, where Z is the fractional
abundance of heavy elements \citep{yi}. See Fig.~9 of \citet{rey07} for an example of how the shape
of this kind of HB-morphology isochrones can be changed by the assumption of a significant spread in
He abundance within the HB population. In our case, the presence of a population of He-rich stars
would prevent the isochrones from reaching SMI values near zero even at high metallicities.}.
We stress that, for the same reasons mentioned above for the age scale, this should be considered just as a rough indication of the effect Y on SMI. It is also important to keep in mind that in our current view of globular clusters, a previously usual concept like ``the He abundance of a given cluster'' is no more valid, since all GCs presenting a spread in light elements are expected to present a spread also in Y (see Sect.~\ref{intro}). Hence each of them is characterized, in this respect, by {\em the distribution of He abundances of its member stars}. Different distributions of Y would lead to differences in HB morphology and, consequently, in SMI. However both the average abundance and the size of the spread (as well as the {\em shape} of the distribution) would concur in determining the final SMI value of a given cluster. Hence any difference in SMI caused by helium, should trace a difference in the Y distributions of the considered clusters.

Fig.~\ref{f:mir} shows the well known fact that clusters residing in the outer halo (OH, $R_{GC}> 8.0$~kpc) of the MW
tend to lie on an isochrone corresponding to a younger age with respect to clusters in the Inner
Halo \citep[IH, $R_{GC}\le 8.0$~kpc~][]{lee94,mar94,mar95,rey01}. This is especially evident at intermediate metallicity,
since at the extremes of the metallicity range the HB morphology parameters like SMI saturate (i.e.
are not able to discriminate within HB distributions that have only red or only blue stars). The
recent analyses by \citet{ghb}, \citet{dot10}, and, in particular, \citet{dot11} confirm that the
observed difference largely traces actual age differences between the two sub-groups of Galactic 
GCs: while IH clusters are approximately coeval at any metallicity, several OH clusters with
[Fe/H]$\ga-1.5$ are $\approx 1-2$~Gyr younger than their IH counterparts with the same metallicity.

The completely new feature of Fig.~\ref{f:mir} is the location of M31 GCs: {\em most Sample~A
clusters with [Fe/H]$\le$--1.2 lie in a different locus with respect to both IH and 
OH Galactic GCs,  approximately along the isochrone corresponding to $\Delta t=-2.2$~Gyr}.  
Specifically, the clusters lying near (or below) the $\Delta t=-2.2$~Gyr isochrone are: B337, 
MCGC9, MCGC8, MCEC2, MCGC5, MCEC4, B336, B514, MCGC10, MCEC1, MCGC1. 
Therefore, eleven  out of  seventeen Sample~A clusters in the considered metallicity 
range, i.e. more than half the sample, lies near this locus.  
These GCs (all labelled in Fig.~\ref{f:mir})
appear to have significantly redder HB morphology with respect to their Galactic counterparts of the
same metallicity. From a different perspective, one can say that a given SMI value is reached by
these M31 clusters at a lower metallicity by $\approx -0.4$~dex than their Galactic counterparts. 
In the following we will refer to these eleven M31 clusters as {\em Anomalously Red (HB) Sample~A} 
clusters, abbreviated with the acronym ARSA. Note also that, for [Fe/H]$\la-0.8$, {\em no Sample~A cluster lie near the $\Delta t = 0.0$ isochrone}, along which most of the Galactic GCs are clustered. 

Only two Galactic GCs are lying on the $\Delta t=-2.2$~Gyr isochrone as the ARSA clusters, i.e.~ 
NGC~4590 and NGC~7078. Both clusters are very metal-poor ([Fe/H]$\la -2.0$) and are not far from 
the blue saturation limit of the SMI. 
The inspection of the CMD of NGC~7078 reveals that the faint limit imposed by our 
definition of SMI ($V<26.0$ for M31 clusters) cuts out a large number of blue HB stars 
belonging to  the extended blue tail,  
namely $\approx$30\% of the whole HB population. 
Therefore, in this case the SMI does not provide a good parametrization of the HB 
morphology, which is actually  bluer than indicated by the SMI. We note, however, 
that virtually all the blue HB stars excluded from the SMI estimate for NGC~7078  lie within 
$\approx 1.5$~mag of the threshold (similar to the case of NGC~6229, illustrated in 
Fig.~\ref{f:method}).
Fig.~\ref{f:cmd1} shows that none of the ARSA clusters seems to have such a large population of 
blue HB stars in this range of magnitude. So it is very unlikely that the SMI of these clusters can 
be severely affected by this kind of problem. 

On the other hand, no HB star of NGC~4590 lies below the faint limit of the selection, and hence 
the SMI accounts for the whole HB distribution, and its relatively low value traces a real difference 
in HB morphology between this cluster and other GCs of similar metallicity.
This peculiarity is remarked also by \citet{ema12}, who showed that there is a deficiency
of very blue HB stars in this cluster with respect to, for example, NGC~7099. 
Since NGC~4590 appears to be as old as the other metal-poor GCs of the MW, \citet{ema12} 
suggest that the observed difference in HB morphology may be accounted for by a difference 
in He abundance\footnote{We note that the two couples of clusters considered by \citet{ema12}, i.e. NGC4590 vs. NGC7099 and NGC5466 vs. NGC6341, display also a large difference in $M_V$ and central density which may play a role in the chemical enrichment of cluster stars. The correlation between these parameters and the extension of blue tails has been noted and discussed in \citet{fp93} and \citet{alehb}.}.  However, it has to be recalled that the apparent anomalous position of these clusters may also be due to inadequacies of the theoretical models, especially in this low metallicity regime, where the shape of the isocrones may strongly depend on the assumptions on mass-loss \citep[see][for examples and discussion]{z93,marcio09}.

Obviously, the unusual location of an {\em individual} M31 GC in the  SMI vs. [Fe/H] plane may be 
considered as hardly significant given the large uncertainties. However, to appreciate the actual sensitivity of SMI for the best Sample~A clusters it may be interesting to consider the 
cases of MCGC5 and B514. Both clusters have very deep and clean CMDs, they have the same metallicity and they differ by $0.45\pm 0.13$ in SMI, formally a $\approx 3.5\sigma$ difference. The comparison of their CMDs (in Fig.~\ref{f:cmd1}) reveals at a first glance that the HB morphology of B514 is indeed much bluer than that of MCGC5, even if both clusters have stars to the red and to the blue of the SMI B-to-R threshold. The same visual comparison can be made for all the ARSA clusters: the cluster-to-cluster differences measured by SMI corresponds to real differences in HB morphology that, in most cases, can be appreciated simply by looking at the CMDs. Moreover,  Sample~A clusters
clearly display a {\em collective} behaviour that  is different from MW GCs  \citep[and similar to dwarf galaxy satellites of the MW and M31, see][]{yangsara}, thus strongly 
hinting at a real difference in some fundamental physical parameter between the samples.

\subsection{The effect of systematics}

Since the observed difference seems to involve the large majority of Sample~A clusters, it would be
easily explained by a systematic error in one of the relevant parameters, i.e metallicity and 
reddening, or SMI itself.

\subsubsection{SMI} 

The most obvious candidate for a systematic effect is our HB morphology parameter. If large numbers of BHB stars were missed in our computation, the actual SMI would be mere lower limits, as in the case of Sample~B (see Sect.~\ref{samB}). Once taken into account, these hypothetical BHB stars would lead ARSA clusters to move near the MW clusters, at least those residing in the outer halo of the MW. 

However, the inspection of the CMD of ARSA clusters (Fig.~\ref{f:cmd1}) reveals that there is no sign of such a significant additional BHB population down to $\approx 3$ magnitudes below the $V_{HB}$ level.
Thus, any hidden BHB should be made by extreme HB and/or blue hook stars, like the faintest HB stars in NGC2808, i.e. stars that would be excluded from the computation of SMI by definition and cannot be at the origin of the difference between M31 and MW GCs shown in Fig.~\ref{f:mir}. Moreover, the presence of a significant population of such stars after a gap in the HB distribution of $\ga 2$~mag would make the overall morphology of ARSA clusters {\em exceedingly} peculiar, calling for an interpretation as the results discussed here.

\subsubsection{Metallicity}

A systematic underestimate of the metallicity of 
M31 clusters by $\approx 0.4$~dex would move all the ARSA clusters on the same isochrone populated by Galactic OH clusters. 
However, such a large systematic error can be safely excluded. First, the 
metallicity estimates used here have been obtained in Paper-I from the comparison of the observed RGB
with templates of Galactic GCs, and hence the adopted metallicity scale is exactly the same for 
the MW and the M31 samples (except for a possible factor that is discussed below). 
Second, in Paper-I we have compared our [Fe/H] estimates 
with three sets of independent spectroscopic estimates and we found 
that systematic differences are at most 0.1-0.2 dex and in any case consistent with zero 
(see Fig.~6 of Paper-I). 
It is particularly reassuring that the agreement with estimates from FeI lines from high 
resolution spectroscopy by \citet{col09} is excellent
($\Delta [Fe/H]_{CMD-FeI}=-0.04\pm 0.21$).

An implicit assumption in the metallicity scale of Paper-I is that M31 clusters are as enhanced in 
$\alpha$-element abundance (w.r.t the Sun) as their  Galactic counterparts. This assumption
was found to be valid, at least for the handful of bright M31 clusters analyzed by
\citet{col09}. However this is not necessarily true for Sample~A clusters. If these clusters 
had a solar [$\alpha$/Fe] ratio, the comparison with the grid of templates of $\alpha$-enhanced
Galactic GCs would lead to underestimate the actual metallicity. Using the formulae provided
by \citet{fer99} we estimated that in this case the [Fe/H] values of Sample~A clusters should be
increased by +0.2~dex to keep them in the same metallicity scale as the Galactic GCs. 
This would reduce the difference between the M31 and the MW clusters shown in Fig.~\ref{f:mir} 
but it would not be sufficient to cancel it out, since the observed effect is 2 times larger. 
Even if a difference in [$\alpha$/Fe] were responsible for part of the observed effect, 
this would be nevertheless an important difference in a fundamental physical parameter of 
the clusters (i.e. the [$\alpha$/Fe] ratio), revealed thanks to Fig.~\ref{f:mir}.

\subsubsection{Reddening}

The other parameter that can produce a systematic effect is reddening. Like metallicity and 
distance, reddening was estimated simultaneously  in Paper-I by a best-fit of the CMD with the 
grid of MW GC templates, and it has been verified that the best-fit value cannot be changed 
by more than $\pm 0.02$~mag without compromising the overall fit by obtaining  unrealistic 
(or nonsense) values of distance and metallicity. 
Moreover, changing the reddening has little direct effect on the final SMI value (by shifting 
the Blue/Red threshold), while it may have a sizable effect on the estimated metallicity. 
It turns out that to obtain the higher [Fe/H], which would be necessary to reduce 
the difference between M31 and MW clusters in Fig.~\ref{f:mir}, a {\em lower reddening  
must be adopted} with respect to Paper-I. However, the comparison of our
reddening values with three independent sets of estimates demonstrates that, if 
any small systematic effect were there, it would go {\em in the opposite sense}, 
since our E(B-V) are on average a few hundredths of a magnitude smaller than those estimated 
by other studies (see Paper-I for details).  

However, for the sake of completeness, we have repeated, for the ARSA clusters, the whole procedure of metallicity estimate performed in Paper-I and the computation of SMI, but forcing 
lower reddening values by 0.02 mag with respect to the best-fit solutions of Paper-I. 
It turns out that (a) the changes in SMI are negligible ($\le 0.05$), and (b) [Fe/H] values typically decrease by 0.1~dex, with a range between 0.05~dex and 0.25~dex, in agreement with the predictions by \citet[][his Eq.~19]{buz95}.
While this clearly helps in reducing 
the difference between Sample~A and the MW sample, it is not sufficient to close the gap, and 
for this `partial result' we pay a toll:  all the CMD fits obtained with the lower E(B-V) values 
are {\em significantly} worse than what found in Paper-I, and in most cases clearly not acceptable.

In conclusion, it seems very unlikely that a systematic misestimate of [Fe/H] or E(B-V) can 
be the (only) cause of the different behaviour of Sample~A and MW clusters reported 
in Fig~\ref{f:mir}. 

\begin{figure} \includegraphics[width=80mm]{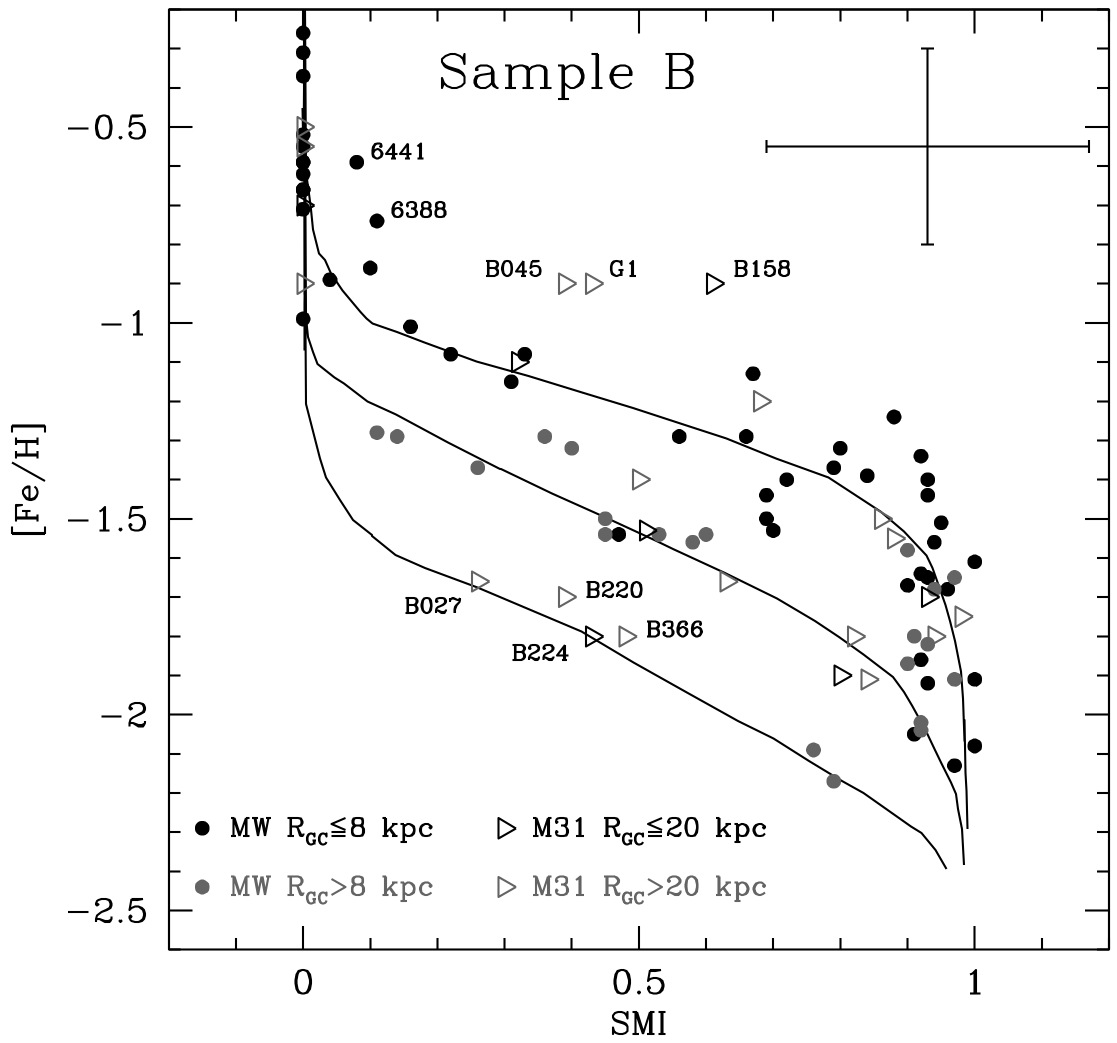} \caption{Same as Fig.~\ref{f:mir} but
for Sample~B M31 clusters. All the estimates should be considered as lower limits, since the bright
limiting magnitude of the available CMD does not allow to reveal a Blue Tail HB component, if present. 
The metal-rich Galactic GCs displaying bimodal HB morphology NGC6441 and NGC6388 are indicated with
labels. Sample~B clusters discussed in the text are also labelled.} 
\label{f:mir_b} 
\end{figure}

\subsection{Sample~B}
\label{samB}

In Fig.~\ref{f:mir_b} we show the distribution of Sample~B clusters in the SMI vs. [Fe/H] plane.
Unfortunately, the uncertainties are too large to draw any firm conclusion, as most of the clusters 
in the interesting range of metallicity have just rough estimates of their SMI. 
In particular, {\em all the SMI estimates for Sample~B clusters should be considered as lower limits}, since the bright
limiting magnitude of the available CMDs does not allow to reveal even the reddest portion of the BHB component, if present. 
However, the comparison with other independent tracers of the HB morphology shown 
in Fig.~4 demonstrates that the estimated SMI may still carry useful information on 
the HB of these clusters.

Taking Fig.~\ref{f:mir_b} at face value, it can be noted that several Sample~B clusters lie on the 
isochrones that are populated also by MW GCs, even at intermediate metallicity. 
Still, there are a few potentially interesting outliers:

\begin{itemize}

\item{} four anomalous clusters lying on the
$\Delta t=-2.2$~Gyr isochrone, as most Sample~A clusters do, namely B027, B220, B224, and B366 \citep[the last has been highlighted and discussed also by][because of its anomalously red $(FUV-V)_0$ colour]{ema12};

\item{}  three metal-rich clusters ([Fe/H]$>-1.0$) which appear to have too blue an HB
morphology for their metallicity (B045, G1 and B158). In particular, G1 is a confirmed case  
of a metal-rich cluster with both red and blue HB stars \citep[see][]{g1}.

\end{itemize}

This feature suggests a similarity with NGC~6441 and NGC~6388, the metal-rich Galactic GCs 
displaying a strong red HB clump as well as a significant blue HB tail, that it is 
currently interpreted as due to a large spread in He abundance \citep{c6441,y6441}. 
Also several bright GCs in M87 are interpreted as hosting a He abundance spread 
\citep[see][their Fig.~9, in particular]{rey07}. 
Therefore, similarly to what is believed to happen in some (likely most) Galactic 
GCs \citep[][and references therein]{ghb,grat12}, also these M31 GCs might 
host (multiple) populations of He-enriched stars, which evolve as Blue HB stars in spite of 
their high metallicity, as already proposed by, e.g. \citet{n2808}. The occurrence of the same phenomenon in some massive and metal-rich M31 clusters has already been suggested with independent 
observational evidence by \citet{rey07,col09,ema12}. The pretty blue $(FUV-V)_0$ colour of many Sample~B clusters, possibly indicative of the presence of unseen BHB/EHB populations, has been already discussed in Sect.~\ref{smifuv}, above. We stress again that Sample~B cluster is mainly composed by luminous/massive (18 on 25 have $M_V\le-9.0$), a range of luminosity where only a few Galactic GCs (and very peculiar ones, like M54, $\omega$~Cen, NGC2419, etc.)  are found.  

It may be interesting to explore the available photometric and spectroscopic databases in
search for further hints on the possible existence of multiple stellar populations in  M31 GCs.  
Such an analysis is in progress, and the results will be reported in a dedicated paper 
(Fusi Pecci et al., in preparation).

\section{Summary and discussion} 
\label{summ}

We performed the first thorough and homogeneous comparison of the HB morphology between M31 and
Milky Way globular clusters, based on HB star counts on cluster CMDs. 
We used a simplified version of the Mironov's index (SMI) that has been demonstrated to
correlate very well with the popular HBR index.
Strict requirements on the
quality of the CMD forced us to limit the core of our analysis to a selected sample of 23 M31 GCs, mostly
located in the outer halo of M31 (Sample~A). 

We find that eleven  of these clusters 
lie on a significant different locus in the metallicity vs HB morphology plane 
with respect to their Galactic counterparts located at any distance from the MW center. 
Having an unusually red HB morphology for their metallicity, we refer to them as 
ARSA clusters (see Sect. \ref{results}).  
The possibility that such a difference arose from systematic errors in some of the involved
parameters has been considered and dismissed as very unlikely.

The most straightforward interpretations of the observed difference
are in terms of age or He abundance differences:  

\begin{enumerate}

\item the ARSA clusters are, on average, $\approx 1$~Gyr younger than the MW Outer
Halo GCs, and $\approx 2$~Gyr younger than the MW inner halo GCs;  or

\item they have a different (internal) distribution of He abundances (Y) than their Galactic
counterparts, i.e. they should have a lower fraction of He-enriched stars 
(see Sect.~\ref{results}).

\end{enumerate}

Hypothesis 1 seems more natural, by analogy with the Milky Way \citep{dot11}. 
In the Galaxy,  slightly younger GCs are generally found to be associated with halo substructures 
and/or relics of disrupting dwarf galaxies \citep{marin}. 
Also most ARSA clusters can be associated to similar substructures in the M31
halo \citep{per09,M10}, and hence younger ages would fit as a general characteristics of 
recently accreted clusters from dwarf satellites\footnote{For example, it is interesting to 
note that the ARSA clusters MCGC8 and MCGC9 are both likely associated with the halo 
filamentary substructure called Eastern Arc \citep[Stream D, in the nomenclature by][]{M10}.}. 
The fact that Sample~A clusters are even 
younger than the MW OH clusters may be connected to the more complex and extended recent 
accretion activity that took place in Andromeda \citep[w.r.t. the MW,][]{iba07}. 
In the process of hierarchical assembly of the MW and M31, the most massive (and most metal rich) 
sub-units were likely the first to merge at the bottom of the potential well, the strong 
compression shocks likely leading to a large amount of star (and cluster) formation on 
a short timescale. At later times less massive and more metal-poor fragments (dwarf galaxies) 
were accreted: these were able to form stars over longer timescales \citep{mateo,tolstoy}. 
It can be conceived that also the epoch of GC formation was consequently postponed in these 
galaxies. In this context it is very interesting to note that, in the [Fe/H] vs. HB morphology plane, dwarf galaxy satellites of the MW and M31 lie virtually in the same locus as ARSA clusters
\citep{yangsara}.

Perhaps the simplest way to envisage hypothesis 2 is to consider the possibility that the 
ARSA clusters do not host multiple populations, as most MW  GCs do, or had a less pronounced early-enrichment history, hence do not present 
the sizable He abundance spread which  acts as an {\em internal} second parameter, by pushing 
a fraction of cluster stars to the blue side of the HB. 
It has been suggested that clusters can host multiple populations (and consequently He enrichment) 
only above a certain mass threshold \citep[a few $10^4~M_{\sun}$, 
corresponding to $M_V\approx -5.5$ according to][and references therein]{grat12}. 
Only one Sample~A cluster is below this threshold, while $\approx$~80\% of the sample 
have $M_V\le -7.0$ and $\approx$~65\% are brighter than the typical average of GC distributions \citep[$M_V=-7.5$,][]{bs06}, hence they appear as massive as Galactic clusters where 
light element (and thus presumably He) enrichment has been observed. Moreover, many ARSA clusters present bimodal HB morphology, possibly suggesting that some intrinsic difference among cluster stars is indeed there \citep[but see][for serious warning on simplistic interpretations of complex morphologies]{fp93}. 
The formation environment may also have played a role \citep{bekki11}, and the physical 
conditions that led to a later formation epoch may have also influenced the formation 
path of these clusters. Therefore, it is  also possible that both age and He abundance have 
concurred in producing the observed difference in HB morphology between outer M31 GCs and 
Milky Way clusters at any distance from the Galactic center. 
It is worth noting that the only Galactic GC in our sample that lies on the same isochrone 
as the ARSA clusters, because it has a genuinely redder HB morphology than other clusters of 
the same metallicity (NGC~4590; see Sect.~\ref{results}),  is as old as them. 
Therefore, at least in this case, the difference in HB morphology must be due to some other 
parameter than age \citep[see, e.g.][]{ema12}. 

Finally, the analysis of sample~B clusters provides support to the conclusions by \citet{rey07} and \citet{ema12} that a population of metal-rich GCs with sizable fractions of Blue HB stars is present in M31, possibly including more extreme cases with respect to their Galactic counterparts (i.e. NGC~6388, NGC~6441), since they show larger SMI values. NGC~6388 and NGC~6441  
have a strongly  bimodal HB morphology that is difficult to explain without invoking a significant spread in He abundance among cluster stars \citep{c6441,y6441}.
Also, many Sample~B clusters have $(FUV-V)_0$ colours suggesting that the presence of extended blue HB tails may be more common in M31 clusters than in the Milky Way. However, it must be noted that the majority of sample~B clusters are very bright (72\% brighter than $M_V= -9.0$ and only one cluster with $M_V>-8.0$).
\citet{euge} found that more luminous/massive clusters appear to have wider distributions of light elements abundance: this should lead to wider distributions of Y and, consequently, to more extended blue tails in the HB. Hence, the above result may simply reflect the case that very massive clusters, possibly with strong anti-correlation patterns and complex HB morphologies including extended blue tails (and blue hooks), like $\omega$~Cen or M54, are much more numerous in the Andromeda galaxy than in the Milky Way\footnote{This does not necessarily mean that the {\em fraction} of extended-blue-tail clusters is higher in M31 than in the MW, since M31 host a much larger number of GCs than the MW \citep[$\ga 450$ vs. $\approx 150$;][]{bar01,gal06}.}. Whatever the physical mechanism that led to such UV-enhanced
HB morphology, its better knowledge in the M31 GC environment could certainly 
add important clues to a better understanding of the early evolution of the
whole cluster system around our nearby galaxy companion, in spite of all the 
observing problems associated with its large distance.
Furthermore, on a larger galactic mass scale, this scenario could also shed 
light to the well recognized UV-upturn phenomenon, that in most cases seems to
constrain the evolution of elliptical galaxies \citep{yi11,buz08,buz12}.


\begin{table*}
  \begin{center}
  \caption{M31 clusters parameters}\label{Tab1}
  \begin{tabular}{lcccccc}
    \hline
Name  & SMI & [Fe/H] & E(B-V) & $R_p$  & $R_{GC}$  &Sample\\
      &     &        &        & [kpc]  & [kpc]     &    \\
     \hline
  B008 &0.00$\pm$0.10  &  -1.00 & 0.07 &  5.6 & 12.1 &A \\
  B292 &0.91$\pm$0.25  &  -1.90 & 0.15 & 16.7 & 33.4 &A \\
  B298 &0.76$\pm$0.22  &  -1.80 & 0.09 & 13.9 & 28.8 &A \\
  B336 &0.55$\pm$0.26  &  -1.90 & 0.10 & 12.7 & 31.5 &A \\
  B337 &0.00$\pm$0.10  &  -1.30 & 0.06 & 13.4 & 15.0 &A \\
  B350 &0.83$\pm$0.31  &  -1.80 & 0.11 & 11.4 & 15.7 &A \\
  B407 &0.00$\pm$0.10  &  -0.40 & 0.10 & 19.4 & 30.9 &A \\
  B514 &0.61$\pm$0.12  &  -1.91 & 0.09 & 54.0 & 58.8 &A \\
  B531 &0.00$\pm$0.10  &  -0.40 & 0.14 & 16.8 & 35.2 &A \\
 MCGC1 &0.78$\pm$0.22  &  -2.15 & 0.12 & 45.3 & 46.9 &A \\
 MCGC2 &0.77$\pm$0.26  &  -1.90 & 0.10 & 32.6 & 52.0 &A \\
 MCGC3 &0.91$\pm$0.24  &  -1.90 & 0.10 & 31.1 & 39.1 &A \\
 MCGC5 &0.16$\pm$0.05  &  -1.90 & 0.11 & 76.9 & 76.9 &A \\
 MCGC7 &0.00$\pm$0.10  &  -0.70 & 0.06 & 17.7 & 75.7 &A \\
 MCGC8 &0.14$\pm$0.08  &  -1.53 & 0.09 & 36.2 & 54.4 &A \\
 MCGC9 &0.09$\pm$0.06  &  -1.40 & 0.16 & 38.0 & 68.2 &A \\
MCGC10 &0.66$\pm$0.18  &  -1.90 & 0.09 & 98.1 &104.0 &A \\
 MCEC1 &0.71$\pm$0.14  &  -1.91 & 0.10 & 13.0 & 25.2 &A \\
 MCEC2 &0.11$\pm$0.05  &  -1.75 & 0.13 & 35.9 & 48.9 &A \\
 MCEC3 &0.84$\pm$0.18  &  -1.91 & 0.09 & 13.7 & 13.7 &A \\
 MCEC4 &0.47$\pm$0.18  &  -1.78 & 0.11 & 58.7 & 65.4 &A \\
  B384 &0.00$\pm$0.10  &  -0.50 & 0.04 & 16.0 & 21.6 &A \\
  B468 &0.00$\pm$0.10  &  -0.70 & 0.06 & 19.7 & 22.5 &A \\
  &&&&&&\\
 B006  &0.00$\pm$0.10  &  -0.55 & 0.08 &  6.3 & 51.4 &B \\
 B010  &0.82$\pm$0.26  &  -1.80 & 0.16 &  5.6 & 41.5 &B \\
 B012  &0.94$\pm$0.29  &  -1.80 & 0.11 &  5.6 & 38.2 &B \\
 B023  &0.00$\pm$0.10  &  -0.90 & 0.28 &  4.3 & 54.5 &B \\
 B027  &0.26$\pm$0.15  &  -1.66 & 0.18 &  5.9 & 36.6 &B \\
 B045  &0.39$\pm$0.26  &  -0.90 & 0.16 &  4.8 & 47.5 &B \\
 B058  &0.50$\pm$0.33  &  -1.40 & 0.11 &  6.8 & 25.2 &B \\
 B088  &0.80$\pm$0.21  &  -1.90 & 0.38 &  3.7 &  8.0 &B \\
 B158  &0.61$\pm$0.39  &  -0.90 & 0.09 &  2.3 &  2.3 &B \\
 B220  &0.39$\pm$0.20  &  -1.70 & 0.06 &  5.1 & 22.0 &B \\
 B224  &0.43$\pm$0.19  &  -1.80 & 0.07 &  5.1 & 11.8 &B \\
 B225  &0.00$\pm$0.10  &  -0.50 & 0.05 &  4.6 & 47.5 &B \\
 B233  &0.51$\pm$0.32  &  -1.53 & 0.10 &  7.9 &  8.7 &B \\
 B240  &0.63$\pm$0.37  &  -1.66 & 0.14 &  7.1 & 28.5 &B \\
 B293  &0.93$\pm$0.21  &  -1.70 & 0.12 & 16.8 & 18.3 &B \\
 B311  &0.98$\pm$0.33  &  -1.75 & 0.25 & 12.8 & 39.8 &B \\
 B338  &0.68$\pm$0.20  &  -1.20 & 0.04 & 10.0 & 35.8 &B \\
 B343  &0.86$\pm$0.26  &  -1.50 & 0.10 & 14.4 & 71.6 &B \\
 B358  &0.84$\pm$0.16  &  -1.91 & 0.05 & 19.4 & 76.6 &B \\
 B366  &0.48$\pm$0.11  &  -1.80 & 0.09 & 11.8 & 41.6 &B \\
 B379  &0.00$\pm$0.10  &  -0.50 & 0.13 & 11.1 & 58.6 &B \\
 B386  &0.32$\pm$0.19  &  -1.10 & 0.04 & 13.8 & 19.9 &B \\
 B405  &0.88$\pm$0.20  &  -1.55 & 0.08 & 17.9 & 68.7 &B \\
 G001  &0.43$\pm$0.09  &  -0.90 & 0.04 & 33.9 & 61.9 &B \\
 B255D &0.00$\pm$0.10  &  -0.70 & 0.14 & 12.3 & 18.5 &B \\
\hline
\end{tabular}
\tablefoot{The reported uncertainties on SMI are formal propagated Poisson errors on star counts. 
We arbitrarily assigned an uncertainty of 0.1 in SMI to clusters having only red HB stars. 
Note that SMI values can be (roughly) converted into HBR, for comparison with other samples, by inverting Eq.~\ref{presto}, i.e $HBR=2SMI-1.0$. Distance moduli, [Fe/H], and E(B-V) are taken from Paper-I. $R_p$ is obtained adopting the mean distance modulus from Paper-I for all 
the clusters, $(m-M)_0=24.42$. The three-dimensional distance from the center of the galaxy, $R_{GC}$, is obtained
adopting the individual distance moduli from Paper-I and $(m-M)_0=24.42$ for the center of M31.} 
\end{center}
\end{table*}

\begin{acknowledgements}
We acknowledge the financial
support to this research by INAF through the PRIN-INAF 2009 grant CRA 1.06.12.10 (PI: R. Gratton)
and by ASI through contracts COFIS ASI-INAF 
I/016/07/0 and ASI-INAF I/009/10/0. We are grateful to Marcio Catelan and Emanuele Dalessandro 
for useful suggestions and discussion. Emanuele Dalessandro kindly shared their data with 
us before publication.

\end{acknowledgements}

\bibliographystyle{apj}


\end{document}